\documentclass[
    ,final            % use final for the camera ready runs
  ]
  {aipproc}

\layoutstyle{6x9}

\begin{document}

\title{A Broader Perspective on the GRB-SN Connection}


\author{Alicia M. Soderberg}{
  address={Division of Physics, Mathematics and Astronomy,105-24, California Institute of Technology, Pasadena, CA 91125}
}

\begin{abstract}
Over the last few years our understanding of local Type Ibc supernovae
and their connection to long-duration gamma-ray bursts has been
revolutionized.  Recent discoveries have shown that the emerging
picture for core-collapse explosions is one of diversity.  Compiling
data from our dedicated radio survey of SNe Ibc and our comprehensive
HST survey of GRB-SNe together with ground-based follow-up campaigns,
I review our current understanding of the GRB-SN connection.  In
particular, I compare local SNe Ibc with GRB-SNe based on the
following criteria: (1) the distribution of optical peak magnitudes
which serve as a proxy for the mass of $^{56}$Ni produced in the
explosion, (2) radio luminosity at early time (few days to weeks)
which provides a measure of the energy coupled to on-axis relativistic
ejecta, and (3) radio luminosity at late time (several years) which 
constrains the emission from GRB jets initially directed away
from our line-of-sight.  By focusing on these three points, I will
describe the complex picture of stellar death that is emerging.
\end{abstract}

\maketitle

%%%%%%%%%%%%%%%%%%%%%%%%%%%%%%%%%%%%%%%%%%%%
%% MAINMATTER
%%%%%%%%%%%%%%%%%%%%%%%%%%%%%%%%%%%%%%%%%%%%

\section{Introduction}

Twenty years have passed since the class of Type Ibc
supernovae (SNe Ibc) was initially recognized as a distinct population
of core-collapse explosions \cite{emn+85,wl85,f97}.  Their lack of
homogeneity and low event rate, ($\sim 10\%$ of locally discovered
SNe), did not motivate focused observational programs.

In 1998, however, SNe Ibc enjoyed an explosion of new-found interest
thanks to the discovery of Type Ic SN\,1998bw ($d\approx 36$ Mpc)
within the {\it BeppoSAX} localization error box of gamma-ray burst,
GRB\,980425 \cite{gvv+98,paa+00}.  While the $\gamma-$ray energy
release, $E_{\gamma}$, of GRB\,980425 was a factor of $10^{4}$ below
that of GRBs, SN\,1998bw was (and still remains) the most luminous
radio SN ever observed \cite{kfw+98}.  Two unusual features were noted
from the radio data: significant energy ($E_{\rm radio} \sim
10^{49}\,$erg) coupled to mildly relativistic (Lorentz factor,
$\Gamma\sim 3$) ejecta and evidence for episodic energy injection
\cite{kfw+98,lc99}.  Moreover, the bright optical emission required
production of $\sim 0.5\,M_\odot$ $^{56}$Ni, comparable to that inferred
for Type Ia supernovae while the broad absorption lines (indicative of
photospheric velocities above 30,000 km~s$^{-1}$) implied a total
kinetic energy of $3\times 10^{52}$ erg \cite{i+98,w+99}.

These observations have been interpreted under the framework of the
``collapsar model'' (e.g. \cite{m+01}) in which a central
engine (accreting black hole) plays a significant role in
exploding the star. 

\section{The GRB-SN Connection: An Overview}

In the seven years since the discovery of SN\,1998bw/GRB\,980425,
about a dozen SNe Ibc have been reported in association with GRBs, all
at $z\ge 0.1$ (see \cite{zkh04,skp+06} for recent compilations). Of
these associations, three were unambiguously confirmed through
spectroscopic identification of SN features (GRB\,030329
\cite{mgs+03}; GRB\,031203 \cite{mtc+04}; XRF\,020903 \cite{skf+05})
which were observed to be unusually broad and similar to those seen in
SN\,1998bw.  The majority of GRB-SN associations are inferred based on
the emergence of a red ``bump'' in the afterglow light-curves
approximately $20(1+z)$ days after the explosion and attributed to a
thermal supernova component.  These observations imply that at least
some SNe Ibc are powered by a central engine.

At the same time, several broad-lined supernovae have been discovered
locally and are currently estimated to represent $\sim 5\%$ of the
Type Ibc population \cite{pmn+04}.  Given their spectral similarity to
SN\,1998bw and GRB-SNe, it has been argued that broad-lined SNe
Ibc can be used as signposts for GRBs, even in the absence of observed
gamma-ray emission (e.g. \cite{mkm+05}).

The question has thus become, what is the connection (if any)
between the engine-driven GRB-associated SNe and local SNe
Ibc?  Here I present optical and radio observations for these two
samples in an effort to address this question and to offer a broader
perspective on the GRB-SN connection.

\begin{figure}
  \includegraphics[height=.5\textheight]{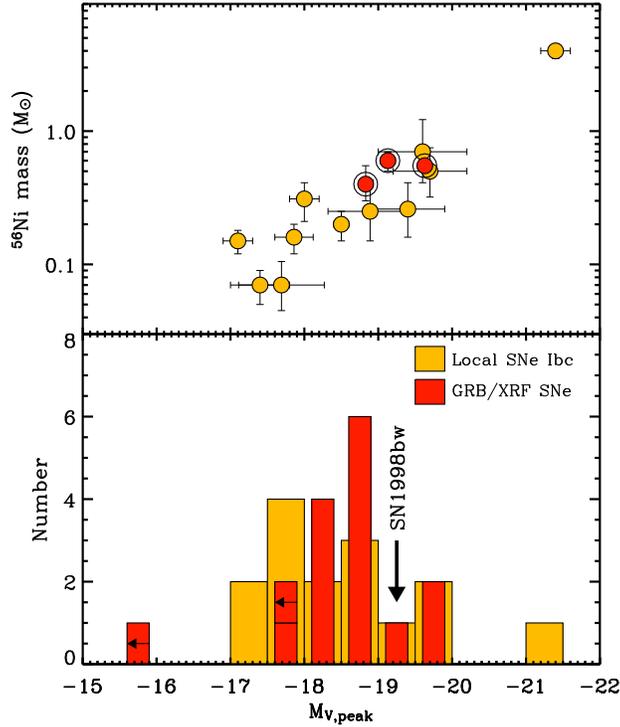}
  \caption{The compilation of peak optical magnitudes and $^{56}$Ni
    mass estimates for GRB-SNe and local SNe Ibc from Soderberg {\it
    et al.} (2006a) has been extended to included
    SN\,2003lw/GRB\,031203 (Mazzali {\it et al.}, in prep) and local
    SNe 2003jd \cite{mkm+05} and 2005bf \cite{ttn+05}.  The
    distributions for GRB-SNe and local SNe Ibc show significant
    overlap.  A K-S test shows a 53\% probability that the two samples
    are drawn from the same parent population of SNe.}
\end{figure}

\section{An Optical Perspective on the GRB-SN Connection}

In Figure~1 I compare the peak optical magnitudes and $^{56}$Ni mass
estimates for GRB-associated SNe and local SNe Ibc.  The figure
clearly shows that $^{56}$Ni mass scales with peak optical luminosity.
As a result, M$_V$ can be used as a proxy for the synthesized
$^{56}$Ni mass in the cases where estimates are not
available.  Several striking conclusions can be drawn directly from
this compilation:

\begin{itemize}
\item
{\bf SN\,1998bw is not the most luminous event of either sample.}  In fact
several local SNe Ibc and GRB-associated SNe are actually
brighter.  This emphasizes the fact that not all GRB-SNe are like
SN\,1998bw.  

\smallskip

\item
{\bf The distributions of local and GRB-associated SNe show significant
overlap.}  We conclude that GRB-associated SNe are not necessarily more
luminous nor do they produce more $^{56}$Ni than local SNe Ibc.  In
fact, a K-S test on the two data samples shows a $53\%$ probability
that the two have been taken from the same parent population of
events.  This may indicate a similar $^{56}$Ni production mechanism
for both samples and thus imposes significant constraints on
progenitor models.

\smallskip

\item
{\bf SNe Ibc with broad optical absorption lines are
  not more luminous than other events.}  In fact, they display a range
of optical luminosities comparable to the spread observed for both the
local and GRB-SN samples.  This emphasizes that broad-optical
absorption lines cannot be used as a proxy for a large $^{56}$Ni mass.
\end{itemize}

These four points illustrate the fact that optical observations cannot
be used to distinguish the class of GRB-SNe from the local SNe Ibc.

\section{A Radio Perspective of the GRB-SN Connection}

Radio observations offer a better way to distinguish between GRB-SNe
and local SNe Ibc since they provide the best calorimetry of the
explosion.  Radio emission from SNe Ibc is produced by the dynamical
interaction of the fastest ejecta with the circumstellar medium
\cite{c98}, in much the same way that GRB afterglows are
produced.  As the ejecta sweep up and shock the surrounding medium
they produce synchrotron emission with a spectral peak near the radio
band on a timescale of days to weeks.  The emission is brightest for
SNe with copious energy coupled to (mildly) relativistic ejecta, as in
the case of SN\,1998bw.  Radio observations are therefore unique in
that they provide a measure of the speed and energy of the fastest
ejecta produced in the explosion.

Motivated thus, since 1999 we have been monitoring local SNe Ibc with
the Very Large Array on a timescale of days to years after the
explosion.  Early observations are used to probe on-axis ejecta
components while late-time data constrain components that were
initially directed away from our line-of-sight.  This six year effort
has resulted in several key advances in our understanding of the
GRB-SN connection.

\begin{figure}
  \includegraphics[height=.4\textheight]{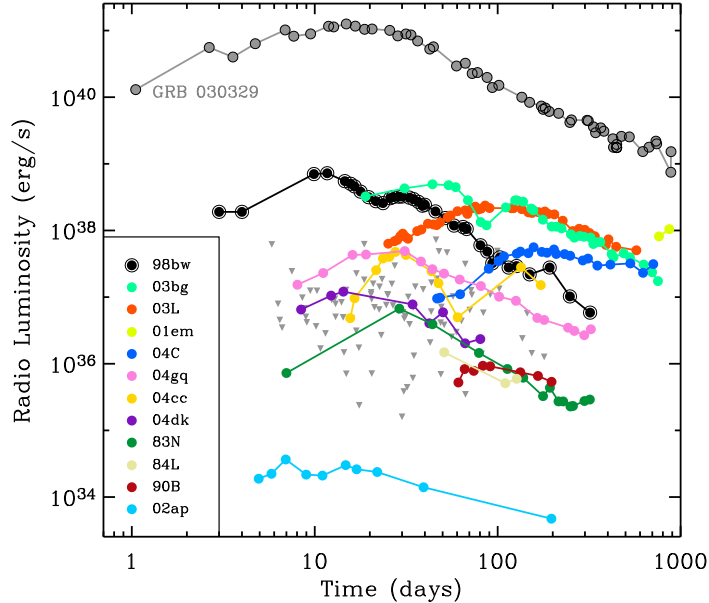}
  \caption{A compilation of SNe Ibc radio observations resulting
    mainly from our dedicated VLA survey (Soderberg {\it et al.} in
    prep).  Upper limits are shown as inverted grey triangles.  All
    detections have been studied as part of our VLA program with the
    exception of SNe 1990B, 1984L, 1983N, 2001em and 1998bw which were
    taken from the literature \cite{vsw+93,psw86,spw84,s+05,kfw+98}.
    For comparison we show the radio light-curve of GRB\,030329 which
    had a radio luminosity typical of long duration GRBs
    \cite{bkp+03}.  These radio data show that there is a clear
    distinction between local SNe Ibc and GRB-SN explosions.}
\end{figure}

\subsection{Early-Time Observations}

First, through our extensive sample of 146 local SNe Ibc, we now know
that only $10\%$ have detectable radio emission on a timescale of a
few days to years (Figures~2 and 3).  These events typically peak in
the radio band several weeks after the explosion with average
luminosities a factor of $10^2$ times fainter than SN\,1998bw on a
comparable timescale.  We compare the optical and radio properties for
this sample of SNe Ibc and find no strong correlations.  In
particular, broad-lined SNe Ibc are {\bf not} more radio luminous than
the rest of the sample and can be significantly fainter
(e.g. SN\,2002ap, \cite{bkc02}; SN\,2003jd, \cite{snb+06}).  We conclude
that radio bright SN\,1998bw-like events are rare: less than $2\%$ of
the local population (\cite{bkf+03}; Soderberg {\it et al.}, in prep).

Next, we find a clear distinction between local SN Ibc explosions
and cosmological GRBs. As clearly shown in Figure~2, GRB-SN explosions
are a factor of about $10^4$ times more radio luminous than typical
SNe Ibc.  This is attributed to the fact that GRB-SN explosions couple
the bulk of their energy to highly relativistic ejecta while SNe
Ibc couple a relatively tiny fraction.  These results strongly suggest
that if central engines power the majority of local SNe Ibc, the
engines must be weaker than those of GRB-SN explosions.

\begin{figure}
  \includegraphics[height=.4\textheight]{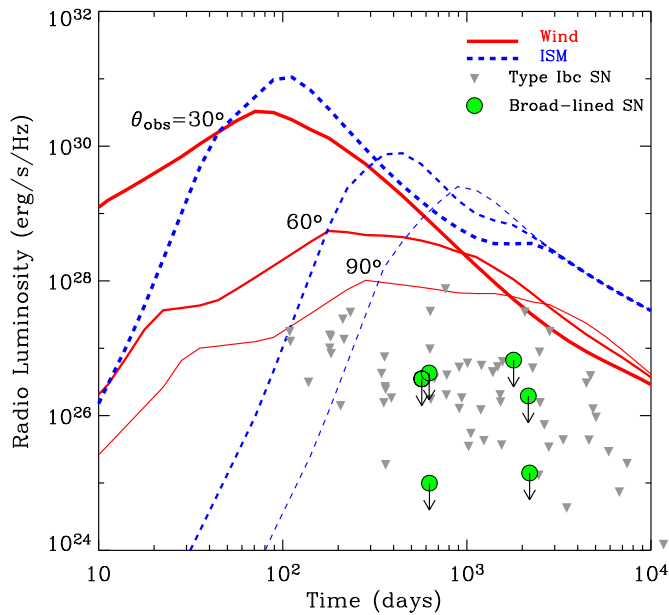}
  \caption{A compilation of late-time radio data for 68 local SNe Ibc
    as published in Soderberg {\it et al.} (2006b).  Upper limits are
    shown as inverted grey triangles and we emphasize the limits for
    broad-lined SNe with green circles/arrows. For comparison we show
    the predicted radio light-curves for a typical GRB afterglow
    ($E=10^{51}$ erg, $n=A_*=1$, $\theta_{\rm jet}=5^{\circ}$)
    observed 30, 60 and 90 degrees away from the collimation axis
    expanding into a constant density medium (dashed blue lines) and a
    stellar wind environment (solid red lines).  Nearly all of the
    upper limits are fainter than the predicted emission from a GRB, 
    even viewed $90^{\circ}$ off-axis.  This holds in particular for
    the broad-lined SNe where we rule out the scenario that every
    broad-lined SN Ibc harbors a GRB.}
\end{figure}

\subsection{Late-Time Observations}

Finally, we use our late-time data obtained between 1 to 30 years
after the explosion to search for evidence of relativistic GRB jets
that were initially directed away from our line-of-sight.  As the jets
sweep up material and decelerate they spread sideways, eventually
intersecting our viewing angle \cite{pac01,wax04}.  At this point the
GRB afterglow emission becomes visible and is most easily detected in
the radio band.  In Figure~3 we show that of the 68 events in our
late-time sample, none show evidence for strong radio emission that
can be attributed to an off-axis GRB jet.

We \cite{snb+06} compare these data to the radio luminosities of
cosmological GRBs to limit the fraction of SNe Ibc hosting GRB jets to
$\le 10\%$ (90\% confidence).  This holds in particular for the
broad-lined events: we rule out a scenario in which every broad-lined
Ibc hosts a GRB ($84\%$ confidence).  This result, taken together with
the early-time radio data, reiterates that {\bf broad optical
absorption lines do not imply the presence of relativistic ejecta.}

\section{Conclusions}

We compare the optical and radio properties for local SNe Ibc and
GRB-SNe.  We show that the optical luminosities for GRB-SNe and local
SNe Ibc are comparable and therefore cannot be used to distinguish
between the two samples.  From our comprehensive radio survey of local
SNe, however, we are able to show a clear distinction between
GRB-SNe and SNe Ibc: the radio luminosities of local events are
typically $10^4$ times fainter than those of typical GRB-SNe and
$10^2$ times fainter than SN\,1998bw.  We conclude that GRB-SN
explosions couple the bulk of their energy to highly relativistic
ejecta while local SNe couple a relatively tiny fraction.  Finally, we
use our late-time radio observations to constrain the fraction of SNe
Ibc harboring off-axis GRBs to less than 10\%.  This holds in
particular for the local broad-lined SNe Ibc which have been argued
(based on their spectral similarity to GRB-SNe) to be associated with
off-axis GRBs.  In conclusion we find that while most GRB explosions
have a supernova component, only a small fraction of SNe Ibc are
capable of producing the copious relativistic ejecta characteristic of
gamma-ray bursts.

%%%%%%%%%%%%%%%%%%%%%%%%%%%%%%%%%%%%%%%%%%%%%%%%
%% BACKMATTER
%%%%%%%%%%%%%%%%%%%%%%%%%%%%%%%%%%%%%%%%%%%%%%%%

%%%%%%%%%%%%%%%%%%%%%%%%%%%%%%%%%%%%%%%%%%%%%%%%
%% The bibliography can be prepared using the BibTeX program or
%% manually.
%%
%% The code below assumes that BibTeX is used.  If the bibliography is
%% produced without BibTeX comment out the following lines and see the
%% aipguide.pdf for further information.
%%
%% For your convenience a manually coded example is appended
%% after the \end{document}
%%%%%%%%%%%%%%%%%%%%%%%%%%%%%%%%%%%%%%%%%%%%%%%%

%%%%%%%%%%%%%%%%%%%%%%%%%%%%%%%%%%%%%%%%%%%%%%%%
%% You may have to change the BibTeX style below, depending on your
%% setup or preferences.
%%
%%
%% For The AIP proceedings layouts use either
%%%%%%%%%%%%%%%%%%%%%%%%%%%%%%%%%%%%%%%%%%%%

\bibliographystyle{aipproc}   % if natbib is available
%\bibliographystyle{aipprocl} % if natbib is missing

%%%%%%%%%%%%%%%%%%%%%%%%%%%%%%%%%%%%%%%%%%%
%% You probably want to use your own bibtex database here
%%%%%%%%%%%%%%%%%%%%%%%%%%%%%%%%%%%%%%%%%%%
%\bibliography{sample}

%%%%%%%%%%%%%%%%%%%%%%%%%%%%%%%%%%%%%%%%%%%
%% Just a reminder that you may have to run bibtex
%% All of it up to \end{document} can be removed
%% if you don't like the warning.
%%%%%%%%%%%%%%%%%%%%%%%%%%%%%%%%%%%%%%%%%%%
%\IfFileExists{\jobname.bbl}{}
% {\typeout{}
%  \typeout{******************************************}
%  \typeout{** Please run "bibtex \jobname" to optain}
%  \typeout{** the bibliography and then re-run LaTeX}
%  \typeout{** twice to fix the references!}
%  \typeout{******************************************}
%  \typeout{}
% }

%\end{document}

%%%%%%%%%%%%%%%%%%%%%%%%%%%%%%%%%%%%%%%%%%%
%% The following lines show an example how to produce a bibliography
%% without the help of the BibTeX program. This could be used instead
%% of the above.
%%%%%%%%%%%%%%%%%%%%%%%%%%%%%%%%%%%%%%%%%%%

\end{document}